\documentstyle[12pt]{ioplppt}          % use this for preprint style
\begin{document}
\title{Initial Hypersurface Formulation:
Hamilton-Jacobi Theory for Strongly Coupled Gravitational Systems}
\author{D.S. Salopek \ftnote{1}{E-mail: dsscosmos@yahoo.com}
\ftnote{2}{Current mailing address: Space-Time Institute, 
14915-105 Avenue, Edmonton, Alberta, Canada T5P 4M2} }
\address{Department of Physics \& Astronomy\\
6224 Agricultural Road, University of British Columbia\\
Vancouver, Canada V6T 1Z1\\}
\begin{abstract}
Strongly coupled gravitational systems describe Einstein gravity and matter
in the limit that Newton's constant $G$ is assumed to be very large.
The {\it nonlinear} evolution of these systems may be solved analytically
in the classical and semiclassical limits
by employing a Green function analysis. Using functional methods in a
Hamilton-Jacobi setting, one may compute the generating
functional (`the phase of the wavefunctional') which satisfies both
the energy constraint and the momentum constraint. Previous results are
extended to encompass
the imposition of an arbitrary initial hypersurface. A Lagrange
multiplier in the generating functional restricts the initial fields,
and also allows one to formulate the energy constraint on the initial
hypersurface. Classical evolution follows as a result of minimizing
the generating functional with respect to the initial fields.
Examples are given describing Einstein gravity interacting
with either a dust field and/or a scalar field. Green functions are
explicitly determined for (1) gravity, dust, a scalar field and
a cosmological constant and (2) gravity and a scalar field interacting
with an exponential potential.  This formalism is useful in solving problems
of cosmology and of gravitational collapse. \\

\end{abstract}

In the limit that Newton's constant $G$ approaches infinity, the
mathematical equations describing Einstein gravity coupled to matter simplify
dramatically. One may safely drop second order spatial gradients
while first order spatial gradients are retained.
Using Hamilton-Jacobi (HJ) theory and the
{\it semiclassical Green function method},
a preceding paper (Salopek 1998)
demonstrated how to construct a general class of  solutions
for matter and gravity in the strong coupling limit. Here,
those powerful methods will be extended to formulate the
initial hypersurface problem.

Strongly coupled gravitational systems are useful in
studies of cosmology and of gravitational collapse.
In cosmology, strongly coupled gravity has been used to describe the evolution
of `long-wavelength fields' arising, for example, from inflation
which then serve
as the primordial initial conditions for structure formation. The wavelength
of these fields exceeds the local value of the Hubble radius, and different
spatial points are no longer in causal contact
(Salopek 1991, Salopek and Bond 1990 and Salopek 1998).
A strongly coupled expansion also appears
in string theory formulations of cosmology (Veneziano 1997).
In problems of gravitational collapse, strongly coupled gravity
describes `velocity-dominated' evolution where the gravitational potential
terms which contain spatial gradients may be neglected as space collapses
rapidly into a singularity.
Numerical studies (Berger 1998, Berger and Moncrief 1993,
Hern and Stewart 1998) indicate that many singularities that appear in general
relativity are velocity-dominated. In addition, Hern and Stewart (1998)
have shown numerically that the evolution of a certain class of Gowdy models
may mimic that of velocity-dominated models
even when spatial derivative terms are not negligible.

The two essential ideas behind generating general solutions to strongly coupled
gravity and matter are: (1) computing the semiclassical Green function
solution of the energy constraint
and (2) convolving the Green function with an arbitrary, gauge-invariant,
initial state. In this paper, I will
generalize the Hamilton-Jacobi formalism for strongly coupled gravity and
matter to encompass the situation where one specifies the fields on some
arbitrary initial hypersurface in superspace.

Hamilton-Jacobi theory for general relativity is useful for two
very important reasons:

(1) It can be used to formulate a primitive theory of quantum gravity.
    At present, there are many problems in solving the Wheeler-DeWitt
    equation (DeWitt 1967) which is the canonical equation for quantum
    gravity.
    The worst is perhaps the problem of infinities. This issue
    may be temporarily postponed by expanding the wavefunctional in powers
    of $\hbar$. At the lowest order, one must solve the Hamilton-Jacobi
    equation for general relativity which is mathematically self-consistent
    and does not require renormalization. This paper develops tools that
    allow one to solve the HJ equation in the strongly coupled limit.
    At higher order in $\hbar$, infinities would certainly appear and
    they would have to treated at some later date presumably using some
    new ideas. However, an expansion
    in powers of $\hbar$ has been very successful in field theory, and
    one strongly suspects that it will be useful in the gravitational
    context although much work remains to be done. (Future prospects for a
    quantum description of cosmology and gravity have been discussed by
    Hartle 1997.)

(2) It yields a covariant formulation of the gravitational field.
    Since its conception in the 1960's, it has been known that the
    Wheeler-DeWitt equation
    refers to the 3-metric $\gamma_{ab}$, but that it does not depend on the
    lapse $N$ nor the shift $N_i$. Quite often,
    this leads to problems in interpreting various approximate solutions.
    In a semiclassical setting, the Hamilton-Jacobi equation also does
    not depend on $N$ and $N_i$. This is actually a blessing because a solution
    of the HJ equation is valid for all choices of the lapse and shift. In this
    sense, it provides a covariant description of gravity.

HJ methods have been particularly useful in solving the
long-wavelength problem of cosmology. In fact, they
yield a transparent and elegant solution for the
{\it nonlinear} evolution of long-wavelength fluctuations in cosmological
models (Salopek 1991, Salopek and Bond 1990, Salopek 1998).
(They are also being applied to string cosmology by Saygili 1998).
In the long-wavelength limit, I feel that attempts to include nonlinear
effects using higher order perturbation methods by
Abramo {\it et al} (1997) are actually more difficult to apply and interpret,
and they tend to obscure the fundamental simplicity of the long-wavelength
problem: different spatial points evolve as homogeneous and independent
universes. (Unruh 1998 has discussed the results of
Abramo {\it et al} 1997 in more detail.)   HJ methods
are very powerful because they allow one to glue these independent spatial
points using the momentum constraint to form one universe. This feature is
crucial if one wishes to compute
higher order terms in the strong coupling expansion (Parry, Salopek and
Stewart 1994). Nonlinear long-wavelength fields also figure prominently
in {\it stochastic inflation}  (Vilenkin 1983, Starobinski 1986,
Salopek and Bond 1991, Linde {\it et al} 1994, Vilenkin 1998). There, the
probability distribution for long-wavelength fields is governed by a
diffusion-type equation.

Other applications of HJ theory to cosmology include:
(1) the construction of inflation models which produce non-Gaussian
fluctuations (Salopek 1992) ---these models are still of observational
interest (Fan and Bardeen 1992, Moscardini {\it et al} 1993);
(2) a relativistic formulation of the Zel'dovich approximation
which describes how nonlinear pancake structures form in the distribution of
galaxies (Croudace {\it et al} 1994, Salopek {\it et al} 1994);
(3) a covariant computation of density perturbations from inflation
with one or two scalar fields (Salopek and Stewart 1995, Salopek 1995).

In classic textbooks (Lanczos 1970,  Goldstein 1981), one usually discusses
Hamilton-Jacobi theory for unconstrained systems
in the context of canonical transformations. However, for general
relativity, the presence of constraints prevents a straightforward application
of these standard methods. For example, one may have expected that one could
apply the method of canonical transformations to gravity after a suitable
phase space-reduction. However, since gravity is a nonlinear theory, this
has not been achieved in practice except in some very special cases
(linear perturbation theory, minisuperspace models, etc.).
Rather, one should apply the powerful Dirac formulation of constrained
systems to the constraints of general relativity (interacting with
a scalar field $\phi$ just to be specific),
\numparts
\begin{equation}
{\cal H}[\pi^{ab}(x), \pi^{\phi}(x), \gamma_{ab}(x), \phi(x)]= 0,
\quad {\rm (energy \; \; constraint)}
\label{constraint1}
\end{equation}
\begin{equation}
{\cal H}_i[\pi^{ab}(x), \pi^{\phi}(x), \gamma_{ab}(x), \phi(x)]= 0,
\quad {\rm (momentum \; \; constraint)}
\label{constraint2}
\end{equation}
\endnumparts
where one replaces the momenta  by functional derivatives of the generating
functional, ${\cal S}[\gamma_{ab}(x), \phi(x)]$:
\begin{equation}
\pi^{ab}(x) = { \delta {\cal S} \over \delta \gamma_{ab}(x) }, \quad
\pi^{\phi}(x) = { \delta {\cal S} \over \delta \phi(x) } \, .
\end{equation}
The two constraint equations (\ref{constraint1}-b)
are self-contained equations for ${\cal S}$, and they
may be taken as the starting point for a HJ formulation of gravity.
If ${\cal S}$ is real, one recovers classical general relativity by
integrating the definition of the momenta:
\numparts
\begin{equation}
\left(\dot\gamma_{ij}-N_{i|j}-N_{j|i}\right)/N =2 \kappa \, \gamma^{-1/2}
\left(2\gamma_{ik}\gamma_{jl}-\gamma_{ij}\gamma_{kl}\right)
{ \delta {\cal S} \over \delta \gamma_{kl} } \, ,
\label{evol.metric}
\end{equation}
\begin{equation}
\left(\dot\phi - N^i\phi_{,i}\right)/N=\kappa \, \gamma^{-1/2}
{\delta {\cal S} \over \delta \phi} \, ,
\label{evol.scalar}
\end{equation}
\endnumparts
where
\begin{equation}
\kappa \equiv 8 \pi G \equiv 8 \pi/ m_P^2 \,
\end{equation}
denotes the gravitational coupling constant.
Actually, HJ methods allows one to integrate the classical
evolution equations (\ref{evol.metric}-b) in an elegant way.

In section 1, I give a brief review of the semiclassical Green function
method of solving strongly coupled gravitational systems as advanced by
Salopek (1998).
In section 2, I describe how to generalize the method by
introducing a Lagrange multiplier in the generating functional to include
the case of specifying the fields on an arbitrary initial hypersurface
in superspace. In earlier work, it was found that
one of the initial fields in the Green function must be set
to zero. A specific choice was made previously, but
here I would like to examine other possibilities.

In section 3, I show how to reduce the number of degrees of freedom appearing
in the energy constraint by four.
In sections 4 and 5, I will construct nontrivial Green functions
for the following situations:

(1) Gravity, dust, a massless scalar field and a cosmological constant,

(2) Gravity and a scalar field with exponential potential.

Section 6 uses specific examples to illustrate the generalized
Green function method as applied to gravity, dust and a scalar field
(without cosmological constant). Section 7 adds a cosmological constant.
In section 8, I consider an advanced example involving gravity and dust.
Here the initial hypersurface is defined in terms of the initial dust field
and the Ricci scalar of the initial 3-metric. Since this system is not
analytically solvable, I expand the generating functional in a series.
This example illustrates quite clearly how the energy constraint restricts
the initial fields.  I also show how to construct the 4-metric using
the HJ formalism. A summary and conclusions are given in section 9.

(In this paper, the matter fields have been rescaled by factors of the
gravitational coupling constant $\kappa$, and consequently $\kappa$ disappears
from most expressions. Consult Salopek (1998) for definitions.)

\section{Review of HJ solutions for strongly coupled gravity and matter}

In a HJ formulation, the energy constraint and the momentum constraint
describing the 3-metric $\gamma_{ab}(x)$ interacting with a dust field
$\chi(x)$ are, respectively, (see, {\it e.g.}, Salopek 1998),
\numparts
\begin{equation}
{\cal H}(x)/ \kappa
= \gamma^{-1/2}\,
\left ( 2 \gamma_{ac} \gamma_{bd} - \gamma_{ab} \gamma_{cd} \right )
{\delta {\cal S} \over \delta \gamma_{ab}}
{\delta {\cal S} \over \delta \gamma_{cd}}
 +  \; {\delta {\cal S} \over \delta \chi} = 0 \, ,
\label{ecs}
\end{equation}
\begin{equation}
{\cal H}_{i}(x)=-2\left(\gamma_{ik} \,
{\delta {\cal S} \over \delta \gamma_{kj}} \right)_{,j} +
{\delta {\cal S} \over \delta \gamma_{kl}} \gamma_{kl,i} +
{ \delta {\cal S} \over \delta \chi } \, \chi_{,i} = 0 \, .
\label{mcs}
\end{equation}
\endnumparts
I will refer to $(\gamma_{ab}(x), \chi(x))$ as the
`original fields'. The {\it semiclassical Green function method}
of solving these two equations is described below.

\subsection{Semiclassical Green function method}

A complete solution of the energy constraint eq.(\ref{ecs})
is one where the number of parameter fields equals the number of
original fields.
The Green function solution given below is a complete solution to the energy
constraint with the additional property that the parameter fields,
$[\gamma^{(0)}_{ab}(x), \chi_{(0)}(x)]$, may be interpreted as
`initial fields' on some initial hypersurface in superspace:
\numparts
\begin{eqnarray}
{\cal G}&& [ \gamma_{ab}(x), \chi(x) |  \gamma^{(0)}_{ab}(x), \chi_{(0)}(x) ]
=  \nonumber \\ && {4 \over 3 } \int d^3 x \;
{1 \over \left ( \chi(x) - \chi_{(0)}(x) \right ) } \;
 \left [ 2 \gamma^{1/4} \; \gamma_{(0)}^{1/4} \;
{\rm cosh} ( \sqrt{3 \over 8} z )
- \gamma^{1/2} - \gamma_{(0)}^{1/2} \right ] \, ,
\label{greenf}
\end{eqnarray}
where
\begin{equation}
z = \; {1 \over 2} \; \sqrt{ {\rm Tr}
\left \{
\ln \left( [h] [h_{(0)}^{-1}] \right )
\ln \left( [h] [h_{(0)}^{-1}] \right )
\right \} } \, ,
\label{defz}
\end{equation}
and $[h]$ and $[h^{(0)}]$ are matrices, each with unit determinant,
whose components are given by
\begin{equation}
[h]_{ab} =\gamma^{-1/3} \gamma_{ab} \, , \quad
[h^{(0)}]_{ab} =\gamma_{(0)}^{-1/3} \gamma^{(0)}_{ab} \, .
\label{conf3m}
\end{equation}
\endnumparts
The integrand of the Green function, eq.(\ref{greenf}), has the familiar
HJ form,
\numparts
\begin{equation}
{1\over 2} { D^2 \over \left ( t - t_{(0)} \right ) }   \, ,
\end{equation}
describing a free particle;
the time $t$ may be identified with the proper time $\chi$
of the dust particle, and
$D^2$ is the distance-squared between `two points',
$(\gamma_{ab}, \gamma^{(0)}_{ab})$, in the space of 3-metrics:
\begin{equation}
D^2 = {8 \over 3} \; \left [ 2 \gamma^{1/4} \; \gamma_{(0)}^{1/4} \;
{\rm cosh} ( \sqrt{3 \over 8} z )
- \gamma^{1/2} - \gamma_{(0)}^{1/2} \right ] \, .
\label{ssdistance}
\end{equation}
\endnumparts
This distance function, eq.(\ref{ssdistance}), was first derived by
DeWitt (1967).

In order to ensure consistency of the approach, it was necessary to
set one of the parameter fields to zero. (This point will be discussed in
further detail in section 2). For example, in Salopek (1998)
the initial dust field was set to zero:
\begin{equation}
\chi_{(0)}(x) = 0 \, .
\label{uniform_dust}
\end{equation}
One then constructs
a general solution to the energy constraint and the momentum constraint
through a superposition over an arbitrary `initial state'
${\cal I}$:
\numparts
\begin{equation}
{\cal S}[\gamma_{ab}(x), \chi(x)] =
{\cal G}[\gamma_{ab}(x), \chi(x)| \; \gamma^{(0)}_{ab}(x),  \chi_{(0)}(x)=0] +
{\cal I}[\gamma^{(0)}_{ab}] \, , \label{cals}
\end{equation}
where the initial 3-metric $\gamma^{(0)}_{ab}(x)$ has been chosen to minimize
${\cal S}$,
\begin{equation}
0 = {\delta {\cal G} \over \delta \gamma^{(0)}_{ab} }  +
{\delta {\cal I} \over \delta \gamma^{(0)}_{ab} } \, .
\label{minimize}
\end{equation}
Here ${\cal I}$ is an arbitrary ``gauge-invariant'' functional of
the initial 3-metric:
\begin{equation}
0= -2\left(\gamma^{(0)}_{ik}
{\delta {\cal I} \over \delta \gamma^{(0)}_{kj} } \right)_{,j} +
{\delta {\cal I} \over \delta \gamma^{(0)}_{kl} }
\gamma^{(0)}_{kl,i} \, .
\label{mcinitial}
\end{equation}
\endnumparts
That is, ${\cal I}[\gamma^{(0)}_{ab}(x)]$, is invariant under
reparametrizations
of the spatial coordinates. After solving the minimization equation
(\ref{minimize}), one solves for
the initial 3-metric as a function of the original fields,
\begin{equation}
\gamma^{(0)}_{ab}(x) \equiv
\gamma^{(0)}_{ab}\left[ \gamma_{ab}(x), \chi(x) \right ] \, ,
\label{inversion}
\end{equation}
One then substitutes this result into eq.(\ref{cals}) to determine the
generating functional ${\cal S}[\gamma_{ab}(x), \chi(x)]$ which
depends solely on the original fields.

The solution (\ref{cals}-c) is an {\it Ansatz}.
It was determined by trial and error.
One can verify that it satisfies the
energy constraint (\ref{ecs}) by computing the functional derivatives
of ${\cal S}$ with respect to $\gamma_{ab}(x)$ and $\chi(x)$.
It also satisfies the momentum constraint (\ref{mcs})
because the initial state ${\cal I}$ was chosen to satisfy the
momentum constraint initially, eq.(\ref{mcinitial}), and the momentum
constraint is preserved upon evolution because its Poisson
bracket with the Hamiltonian vanishes.  Many examples given in Salopek (1998)
demonstrate how to exploit the Green function method and the superposition
principle in practice.

\subsection{Initial hypersurface prescription}

The simplest interpretation of setting the parameter field
$\chi_{(0)}(x)$ to zero,
eq.(\ref{uniform_dust}), is that one is choosing an `initial hypersurface'
where the original dust field vanishes.
However, this choice is not forced upon us. In fact,
in defining an initial hypersurface, one may
set any one scalar function of the parameter fields to zero.
I will refer to this process as the
{\it initial hypersurface prescription}. For example,
if, in addition to a dust field, there is a scalar
field $\phi(x)$ and if $\phi_{(0)}(x)$ denotes the corresponding
initial field, one can instead set
\numparts
\begin{equation}
\phi_{(0)}(x) = 0 \,  .
\label{phi0constraint}
\end{equation}
This choice corresponds to defining an initial hypersurface
where the scalar field vanishes.
In addition, one now assumes that ${\cal I}$ is a gauge-invariant functional,
\begin{equation}
{\cal I} \equiv {\cal I}[\gamma^{(0)}_{ab}(x), \chi_{(0)}(x)] \, ,
\end{equation}
of the remaining parameter fields, $\gamma^{(0)}_{ab}(x)$ and $\chi_{(0)}(x)$.
The initial hypersurface condition can
presumably contain spatial derivatives such as,
\begin{equation}
\phi_{(0)}^2 + \gamma^{(0)}_{ab} \, \phi^{(0)}_{,a} \phi^{(0)}_{,b} = 0 \, ,
\end{equation}
or even the Ricci scalar $R_{(0)}$ associated with the initial 3-metric such
as,
\begin{equation}
\chi_{(0)} - B R_{(0)} = 0 \, , \quad ({\rm B \; is \; a \; constant}) .
\end{equation}
\endnumparts
This last example will be treated in detail in section 8.

In general one should be
able to impose any one constraint on the parameter field
such as eqs.(\ref{phi0constraint}-c) by employing a Lagrange multiplier.
I will develop this idea in the next section.

\section{The Lagrange multiplier method}

Although a Green function solution such as eq.(\ref{greenf}-c) to the
energy constraint (\ref{ecs})
is definitely a useful device, it leads to an inconsistency if one
tries to construct solutions to the energy constraint and the
momentum constraint by superimposing over {\it all}
of the initial fields. This mathematical reason for this inconsistency
is that
the initial  fields are not all independent since they obey the energy
constraint on the initial hypersurface. Apparently, the initial energy
constraint reduces the number of degrees of freedom by one field
per spatial point. In order to avoid this
problem, it was suggested in Salopek (1998) that one arbitrarily set one
of the parameter fields to zero before superimposing over the remaining
parameter fields. As was argued on physical grounds in section 1.2,
this choice defines the initial hypersurface.
In the present section, I will expand on this theme by
introducing a Lagrange multiplier $L(x)$
to impose this reduction in the number of parameter fields.
This simple extension leads to a much wider field of applications for
the HJ formalism.  The role of the initial energy constraint will be
discussed using an example in section 8.

\subsection{Formal HJ solution for gravity, dust and a scalar field}

In this section, I will construct a formal HJ solution to a
strongly coupled system consisting of
gravity, a dust field and a scalar field $\phi(x)$ with potential $V(\phi)$.
This system will be described by the following constraint equations:
\numparts
\begin{eqnarray}
{\cal H}(x)/ \kappa
= && \;
\gamma^{-1/2}\,
\left ( 2 \gamma_{ac} \gamma_{bd} - \gamma_{ab} \gamma_{cd} \right )
{\delta {\cal S} \over \delta \gamma_{ab}}
{\delta {\cal S} \over \delta \gamma_{cd}}
 +  \; {\delta {\cal S} \over \delta \chi}  \nonumber \\
&& + {\gamma^{-1/2} \over 2 }
\left ( {\delta {\cal S} \over \delta \phi} \right )^2
+    \; \gamma^{1/2} \, V(\phi)  \; = 0 \; ,
\label{ecgds}
\end{eqnarray}
\begin{equation}
{\cal H}_i(x) = -2\left(\gamma_{ik} \,
{\delta {\cal S} \over \delta \gamma_{kj}} \right)_{,j} +
{\delta {\cal S} \over \delta \gamma_{kl}} \gamma_{kl,i} +
{ \delta {\cal S} \over \delta \phi } \, \phi_{,i} +
{ \delta {\cal S} \over \delta \chi } \, \chi_{,i}
 = 0 \, .
\label{mcgds}
\end{equation}
\endnumparts
For this case, there
are eight degrees of freedom per spatial point associated with the original
fields, $\gamma_{ab}(x)$, $\phi(x)$ and $\chi(x)$.

An {\it Ansatz} solution to the energy and momentum constraints is
\begin{equation}
{\cal S}[\gamma_{ab}(x), \phi(x), \chi(x)] =
{\cal G} + {\cal I} + {\cal L} \, .
\label{gil}
\end{equation}
I will assume that the Green function ${\cal G}$ satisfies the
energy constraint eq.(\ref{ecgds}) with ${\cal S}$ replaced by ${\cal G}$:
\numparts
\begin{eqnarray}
{\cal H}(x)/ \kappa
= && \;
\gamma^{-1/2}\,
\left ( 2 \gamma_{ac} \gamma_{bd} - \gamma_{ab} \gamma_{cd} \right )
{\delta {\cal G} \over \delta \gamma_{ab}}
{\delta {\cal G} \over \delta \gamma_{cd}}
 +  \; {\delta {\cal G} \over \delta \chi}  \nonumber \\
&& + {\gamma^{-1/2} \over 2 }
\left ( {\delta {\cal G} \over \delta \phi} \right )^2
+    \; \gamma^{1/2} \, V(\phi)  \; = 0 \; .
\label{ecgreen}
\end{eqnarray}
It is a complete solution in that it
depends on eight inhomogeneous parameter fields (initial fields),
$\gamma^{(0)}_{ab}(x)$, $\phi_{(0)}(x)$ and $\chi_{(0)}(x)$:
\begin{equation}
{\cal G} \equiv {\cal G}[ \gamma_{ab}(x), \phi(x),  \chi(x) |
\gamma^{(0)}_{ab}(x), \phi_{(0)}(x), \chi_{(0)}(x) ] \, .
\end{equation}
${\cal I}$, the `generalized initial state', is a functional of all the
parameter fields,
\begin{equation}
{\cal I} \equiv {\cal I}[\gamma^{(0)}_{ab}(x), \phi_{(0)}(x), \chi_{(0)}(x) ]
\,  .
\end{equation}
The new ingredient, ${\cal L}$, is a functional of the parameter fields
which is linear in the Lagrange multiplier $L(x)$,
\begin{equation}
{\cal L} = \int d^3 x \, \gamma_{(0)}^{1/2} \; L(x) \,
f\left [ \gamma^{(0)}_{ab}(x), \phi_{(0)}(x),  \chi_{(0)}(x) \right ] \, .
\label{Lagrange}
\end{equation}
\endnumparts
Here $f$ is some scalar function of the parameter fields,
$\gamma^{(0)}_{ab}(x)$, $\phi_{(0)}(x)$ and $\chi_{(0)}(x)$
which will, in general, contain spatial derivatives.
I will demand that ${\cal I}$ and ${\cal L}$ be invariant
under spatial coordinate reparameterizations of the
parameter fields {\it and} the Lagrange multiplier $L(x)$:
\numparts
\begin{equation}
0= -2\left(\gamma^{(0)}_{ik}
{\delta {\cal I} \over \delta \gamma^{(0)}_{kj} } \right)_{,j} +
{\delta {\cal I} \over \delta \gamma^{(0)}_{kl} } \,\gamma^{(0)}_{kl,i} +
{\delta {\cal I} \over \delta \phi_{(0)}} \, \phi^{(0)}_{,i} +
{\delta {\cal I} \over \delta \chi_{(0)}} \, \chi^{(0)}_{,i}
\label{Isymmetric}
\end{equation}
\begin{equation}
0= -2\left(\gamma^{(0)}_{ik}
{\delta {\cal L} \over \delta \gamma^{(0)}_{kj} } \right)_{,j} +
{\delta {\cal L} \over \delta \gamma^{(0)}_{kl} } \,\gamma^{(0)}_{kl,i} +
{\delta {\cal L} \over \delta \phi_{(0)}} \, \phi^{(0)}_{,i} +
{\delta {\cal L} \over \delta \chi_{(0)}} \, \chi^{(0)}_{,i}
+ {\delta {\cal L} \over \delta L } \, L_{,i}
\label{Lsymmetric}
\end{equation}
\endnumparts

The parameter fields and the Lagrange multiplier are chosen
to minimize the generating functional, eq.(\ref{gil}):
\numparts
\begin{eqnarray}
0 = &&{\delta {\cal S} \over \delta \gamma^{(0)}_{ab}(x) } \, ,
\label{la} \\
0 = &&{\delta {\cal S} \over \delta \phi_{(0)}(x) } \, , \label{lb} \\
0 = &&{\delta {\cal S} \over \delta \chi_{(0)}(x) } \, , \label{lc} \\
0 = &&{\delta {\cal S} \over \delta L(x) } \, . \label{ld}
\end{eqnarray}
\endnumparts
The minimization process leads to a solution of the classical Einstein
equations in the strongly coupled limit.

Variation with respect to the Lagrange multiplier $L(x)$,
eq.(\ref{ld}),
implies that the parameter fields are constrained according to,
\begin{equation}
0 = f\left [ \gamma^{(0)}_{ab}(x), \phi_{(0)}(x),  \chi_{(0)}(x) \right ] \, .
\label{fconstraint}
\end{equation}
This effectively reduces the number of parameter fields by one
in accordance with the initial hypersurface prescription of section 1.2.
The minimization method will be successful if one can
solve eqs.(\ref{la}-d) for the parameter fields and the Lagrange multiplier
in terms of the original fields:
\numparts
\begin{eqnarray}
\gamma^{(0)}_{ab}(x) && \equiv
\gamma^{(0)}_{ab} \left[ \gamma_{ab}(x), \phi(x), \chi(x) \right ] \, ,
\label{param1}\\
\phi_{(0)}(x) && \equiv
\phi_{(0)} \left[ \gamma_{ab}(x), \phi(x), \chi(x) \right ] \, ,
\label{param2}\\
\chi_{(0)}(x) && \equiv
\chi_{(0)} \left[ \gamma_{ab}(x), \phi(x), \chi(x) \right ] \, ,
\label{param3}\\
L(x) && \equiv
L \left[ \gamma_{ab}(x), \phi(x), \chi(x) \right ] \, .
\label{param4}
\end{eqnarray}
\endnumparts
If, for example, there are too many parameter fields, something will go wrong
at this stage.
Substitution of the above into eq.(\ref{gil}) leads to the ultimate goal,
the generating functional ${\cal S}$, which depends only on the original
fields.

\subsection{Justification of {\it Ansatz}}

In order to justify the {\it Ansatz} (\ref{gil}), it is necessary to
compute
the functional derivatives of the generating functional with respect
to the original fields.

Before one applies the minimization prescription,
${\cal S}$ is a functional of the original fields, the parameter fields
and the Lagrange multiplier:
\begin{equation}
{\cal S} \equiv {\cal S} \left [\gamma_{ab}(x), \phi(x), \chi(x) |
\gamma^{(0)}_{ab}(x), \phi_{(0)}(x), \chi_{(0)}(x), L(x) \right ] \, .
\end{equation}
After applying the minimization prescription, the parameter fields
and the Lagrange multiplier become functions of the original fields
through eqs.(\ref{param1}-d).
Hence a functional derivative of ${\cal S}$ with the respect to one of the
original fields, say $\phi(x)$, can be computed by the chain rule:
\numparts
\begin{eqnarray}
{\delta {\cal S} \over \delta \phi(x)}\Bigg |_{\gamma_{ab}, \chi} && =
{\delta {\cal G} \over \delta \phi(x)}\Bigg
|_{\gamma_{ab},  \, \chi,
\, \gamma^{(0)}_{ab}, \, \phi_{(0)}, \, \chi_{(0)} } + \nonumber \\
&&\int d^3y \Bigg [ {\delta {\cal S} \over \delta \gamma^{(0)}_{ab}(y)} \,
{\delta \gamma^{(0)}_{ab}(y) \over \delta \phi(x) } +
{\delta {\cal S} \over \delta \phi_{(0)}(y)} \,
{\delta \phi_{(0)}(y) \over \delta \phi(x) } + \nonumber \\
&& {\delta {\cal S} \over \delta \chi_{(0)}(y)} \,
{\delta \chi_{(0)}(y) \over \delta \phi(x) } +
{\delta {\cal S} \over \delta L(y)} \,
{\delta L(y) \over \delta \phi(x) } \,
\Bigg ] \, .
\label{chain_rule}
\end{eqnarray}
All of the terms in eq.(\ref{chain_rule}) which are integrated over $y$
vanish by virtue of the
minimization conditions, eqs.(\ref{la}-d). Hence a functional
derivative of ${\cal S}$ with respect to $\phi(x)$ coincides with a
functional derivative of ${\cal G}$ with respect to $\phi(x)$:
\begin{equation}
{\delta {\cal S} \over \delta \phi(x)}\Bigg |_{\gamma_{ab}, \chi}  =
{\delta {\cal G} \over \delta \phi(x)}\Bigg
|_{\gamma_{ab},  \, \chi,
\, \gamma^{(0)}_{ab}, \, \phi_{(0)}, \, \chi_{(0)} } \, .
\end{equation}
\endnumparts
This argument also holds for functional derivatives with respect to
the other original fields, $\gamma_{ab}(x)$ and $\chi(x)$.

Hence, ${\cal S}$ given by eq.(\ref{gil}) satisfies the energy constraint
(\ref{ecgds}) because:

\noindent
(1) ${\cal G}$ satisfies the energy constraint (\ref{ecgreen}), and

\noindent
(2) functional derivatives of ${\cal G}$ and ${\cal S}$ are
identical, which was illustrated in the previous paragraph.

Moreover, ${\cal S}$ also satisfies the momentum constraint
(\ref{mcgds}) since I demanded
invariance under reparameterization of the spatial coordinates
of both ${\cal I}$ and ${\cal L}$ through eqs.({\ref{Isymmetric}-b).
Since reparametrization invariance is valid initially,
it is guaranteed to be valid at other times.

Thus, ${\cal S}$ given by eq.(\ref{gil}) is a HJ solution to both the
energy constraint and the momentum constraint {\it provided} one
can solve for the initial fields and the Lagrange multiplier
in terms of the original fields, eqs.(\ref{param1}-d).

\subsubsection{Discussion of Lagrange multiplier method}

I will now demonstrate the physical significance of the
Lagrange multiplier method. The functionals, ${\cal I}$ and ${\cal L}$,
appearing in eq.(\ref{gil}) effectively
define the `initial setting' of the parameter fields.

As was mentioned briefly in section 1.2, different choices of the $f$
function, eq.(\ref{fconstraint}),  specify the initial hypersurface.
For example, if
\numparts
\begin{equation}
f = \chi_{(0)} \, ,
\end{equation}
then variation of the Lagrange multiplier implies that $\chi_{(0)}(x)$
vanishes. After setting  $\chi_{(0)}(x)= 0$ in the
generating functional, eq.(\ref{gil}), one recovers eq.(\ref{cals}-c)
which is the result given in Salopek (1998). There, it was shown by explicit
construction that if the dust field were zero, $\chi(x) = 0$,
then the generating functional ${\cal S}$ and the initial state ${\cal I}$
coincide:
\begin{equation}
{\cal S}[\gamma_{ab}(x), \,  \chi(x)=0 ] =
{\cal I}[\, \gamma^{(0)}_{ab}(x) = \gamma_{ab}(x)] \, .
\label{initialstatechi}
\end{equation}
\endnumparts
In this sense, $\chi(x) = 0$ denotes the initial hypersurface
which was chosen among all possible choices by setting the parameter
field  $\chi_{(0)}(x)= 0$.

If on the other hand,
\numparts
\begin{equation}
f = \phi_{(0)} \, ,
\end{equation}
then the scalar field is uniformly zero on the initial hypersurface:
$\phi(x) = 0$. As was shown in Salopek (1998) for the case of a single
scalar field interacting with gravity, if $\phi(x) = 0$,
the generating functional ${\cal S}$ and initial state ${\cal I}$ coincide:
\begin{equation}
{\cal S}[\gamma_{ab}(x), \, \phi(x)= 0] =
{\cal I}[\, \gamma^{(0)}_{ab}(x) = \gamma_{ab}(x)] \, .
\label{initialstatephi}
\end{equation}
\endnumparts
Once again, setting a parameter field to zero effectively determines the
initial hypersurface in superspace.

The choices for $f$ are limitless and they really depend on the problem
at hand. The reader should be warned, however, that if $f$ depends on
spatial derivatives of the parameter fields, then it may be difficult
in practice to solve for the Lagrange multiplier and the parameter fields
as required by eqs.(\ref{param1}-d). Such a case is discussed in section 8.

The general formalism is now complete. The basic result is encapsulated
in the simple expression for the generating functional, eq.(\ref{gil}):
${\cal S} = {\cal G} + {\cal I} + {\cal L}$.  The Green function
${\cal G}$ effectively describes how the system described by the
original fields
$[\gamma_{ab}(x), \phi(x), \chi(x)]$ evolves from initial fields
$[\gamma^{(0)}_{ab}(x), \phi^{(0)}(x), \chi^{(0)}(x)]$.
${\cal I}$ and ${\cal L}$ describe the initial setting. ${\cal I}$ is
the initial state; ${\cal L}$, the Lagrange multiplier term,
prescribes the initial hypersurface.
Classical evolution follows from the minimization of the generating
functional with respect to the initial fields and the Lagrange multiplier.

\section{Reduced energy constraint}

In sections 4 and 5, I will expand the list of known Green functions to include
(1) gravity, dust, a massless scalar field and a cosmological constant and
(2) gravity and a scalar field with an exponential potential.
Before constructing these solutions, I will describe how to reduce
the energy constraint to a more manageable form.

In searching for the Green function solution to the energy constraint
eq.({\ref{ecgds}) describing gravity, dust, and a scalar field
with potential, one must deal with eight degrees of freedom: two matter
fields $\phi$ and $\chi$, and the six components of the symmetric matrix,
$\gamma_{ab}$. One may think that it is hopeless to solve this equation,
but the presence of symmetries makes the task manageable. One attempts a
solution of the form,
\numparts
\begin{equation}
{\cal G} \equiv {\cal G}[ \phi(x), \chi(x), \alpha(x), z(x) ] \, ,
\label{simplerform}
\end{equation}
where the 3-metric degrees of freedom are parameterized by
$\alpha$ and $z$,
\begin{equation}
\alpha = {1\over 6} \ln \gamma \, , \quad
z = \; {1 \over 2} \; \sqrt{ {\rm Tr}
\left \{
\ln \left( [h] [h_{(0)}^{-1}] \right )
\ln \left( [h] [h_{(0)}^{-1}] \right )
\right \} } \, ,
\end{equation}
and $[h]$ and $[h^{(0)}]$ were defined in eq.(\ref{conf3m}).
The `reduced energy constraint' involves only four degrees of freedom:
\begin{eqnarray}
{\delta {\cal G} \over \delta \chi } &&-
{1 \over 12 } \, e^{-3 \alpha} \,
\left( {\delta {\cal G} \over \delta \alpha } \right )^2
+{1 \over 2 }  \, e^{-3 \alpha} \,
\left( {\delta {\cal G} \over \delta z} \right )^2
+{1 \over 2 }  \, e^{-3 \alpha} \,
\left( {\delta {\cal G} \over \delta \phi } \right )^2
+  e^{3 \alpha} \, V(\phi) = 0 \, ,
\nonumber \\
&& {\rm (reduced \; energy \; constraint).}
\label{ecreduced}
\end{eqnarray}
\endnumparts
Not every solution to the energy constraint has
the form (\ref{simplerform})
but the Green function does. In fact, the Green function solutions given
in sections 4 and 5 are derived from the reduced energy constraint
(\ref{ecreduced}).
It is much easier to work with the reduced equation than the full energy
constraint (\ref{ecgds}) because the former does not contain tensor indices.

\section{Green function solution: gravity, dust, massless scalar field with
cosmological constant}

If the potential $V(\phi) = V_0$ describes a cosmological constant, then
the Green function solution of the energy constraint (\ref{ecgds}) for
gravity, dust and a massless scalar field is:
\numparts
\begin{eqnarray}
&& {\cal G}[ \gamma_{ab}(x), \chi(x), \phi(x) |
\gamma^{(0)}_{ab}(x), \chi_{(0)}(x), \phi_{(0)}(x) ]  =
\int d^3x \,  { 2 H_0 \over {\rm sinh}
\left [ {3 H_0 \over 2} \left ( \chi - \chi_{(0)} \right ) \right ]  }  \,
\nonumber
\\
&&\Bigg \{ 2 \gamma^{1/4} \, \gamma_{(0)}^{1/4} \,
{\rm cosh}\left ( \sqrt{3 \over 8}
\sqrt{ z^2 + \left( \phi - \phi_{(0)} \right )^2 } \right ) \nonumber \\
&& - \left ( \gamma^{1/2} + \gamma_{(0)}^{1/2} \right ) \,
{\rm cosh}
\left [ {3 H_0 \over 2} \left ( \chi - \chi_{(0)} \right ) \right ]
\Bigg \} \, , \label{gdmsc}
\end{eqnarray}
where
\begin{equation}
H_0 = \sqrt{V_0 \over 3 } \, .
\end{equation}
\endnumparts
This solution depends on eight parameter fields,
$\left(\gamma^{(0)}_{ab}(x), \chi_{(0)}(x),\phi_{(0)}(x) \right )$.
I will now discuss special cases of this rich solution.

\subsection{Gravity, dust and cosmological constant}

The Green function solution for gravity, dust and a cosmological
constant ({\it i.e.}, no scalar field) is found by setting
\begin{equation}
\phi - \phi_{(0)} = 0 \,
\label{noscalar}
\end{equation}
in eq.(\ref{gdmsc}).
A complete solution depending on seven parameters,
$\left(\gamma^{(0)}_{ab}(x), \chi_{(0)}(x) \right )$, is:
\begin{eqnarray}
{\cal G}&&[ \gamma_{ab}(x), \chi(x) | \gamma^{(0)}_{ab}(x), \chi_{(0)}(x) ]
  =  \int d^3x \,  { 2 H_0 \over {\rm sinh}
\left [ {3 H_0 \over 2} \left ( \chi - \chi_{(0)} \right ) \right ]  }
\nonumber \,  \\
&&\left \{ 2 \gamma^{1/4} \, \gamma_{(0)}^{1/4} \,
{\rm cosh}\left ( \sqrt{3 \over 8} z \right ) -
\left ( \gamma^{1/2} + \gamma_{(0)}^{1/2} \right ) \,
{\rm cosh}
\left [ {3 H_0 \over 2} \left ( \chi - \chi_{(0)} \right ) \right ]
\right \} \, .
\label{gdcc}
\end{eqnarray}
One can readily verify that the above solution reduces to the case of gravity
and dust, eq.(\ref{greenf}), in the limit
that the cosmological constant vanishes, $H_0 \rightarrow 0$.

In general, it is not permissible to arbitrarily specify
{\it original fields} in the Green function (\ref{gdmsc}), and then
expect that they will
satisfy the energy constraint. However, the solution given in
eq.(\ref{gdcc}) may be derived from eq.(\ref{gdmsc}) using the
following argument. If one
assumes that the functionals, ${\cal I}$ and ${\cal L}$,
in eq.(\ref{gil}) that define the
initial setting are independent of $\phi_{(0)}(x)$,
then eq.(\ref{noscalar}) follows from minimization with respect to
$\phi_{(0)}(x)$. As a result, the scalar field dependence in the Green
function (\ref{gdmsc}) drops out, and one loses this degree of freedom.

\subsection{Gravity, massless scalar field and cosmological constant}

The Green function,
\begin{eqnarray}
{\cal G}[\gamma_{ab}(x), &&\phi(x)| \; \gamma^{(0)}_{ab}(x), \phi_{(0)}(x) ] =
- \sqrt{ 4 V_0 \over 3 } \int d^3 x \; \nonumber \\
&&\left[ \gamma+ \gamma_{(0)} - 2 \gamma^{1/2} \gamma_{(0)}^{1/2}
{\rm cosh}
\left ( \sqrt{3 \over2} \sqrt{ z^2 + (\phi - \phi_{(0)} )^2 } \right )
\right ]^{1/2} \; ,
\label{greenmc}
\end{eqnarray}
for gravity, a massless scalar field and a cosmological
constant ({\it i.e.}, no dust field)
was given in Salopek (1998). The sign preceding the Green function is
arbitrary and I have chosen it to correspond to an expanding universe.
One may derive it from the Green function eq.(\ref{gdmsc})
in the following way: if the functionals,  ${\cal I}$ and ${\cal L}$,
in eq.(\ref{gil}) that define the initial setting are independent of
$\chi_{(0)}(x)$, then minimization of ${\cal S}$ with respect to
$\chi_{(0)}(x)$ implies
\begin{equation}
{\delta {\cal G} \over \delta \chi_{(0)}(x) }= 0 \,
\end{equation}
which yields eq.(\ref{greenmc}). Thus the dust degree of freedom drops out.

\subsection{Gravity, dust and a massless scalar field}

The Green function for  dust, gravity and a massless scalar field
({\it i.e.}, no cosmological constant) is
found from eq.(\ref{gdmsc}) by considering the limit
that the cosmological constant vanishes, $H_0 \rightarrow 0$:
\begin{eqnarray}
{\cal G}[ && \gamma_{ab}(x), \phi(x), \chi(x) |
\gamma^{(0)}_{ab}(x), \phi_{(0)}(x), \chi_{(0)}(x) ]
= {4 \over 3 } \int d^3 x \;
{1 \over \left ( \chi(x) - \chi_{(0)}(x) \right ) } \;  \nonumber \\
&& \left [ 2 \gamma^{1/4} \; \gamma_{(0)}^{1/4} \;
{\rm cosh} \left ( \sqrt{3 \over 8}
\sqrt{ z^2 + \left( \phi - \phi_{(0)} \right )^2 } \right )
- \gamma^{1/2} - \gamma_{(0)}^{1/2} \right ] \, ,
\label{greenff}
\end{eqnarray}

\subsection{Special Case: $\gamma_{(0)} \rightarrow 0$ limit}

Note that in the limit that the determinant of the initial 3-metric
approaches zero, $\gamma_{(0)} \rightarrow 0$, the Green function
for gravity, dust, scalar field and a cosmological constant,
eq.(\ref{gdmsc}), takes on the simple form
\numparts
\begin{equation}
{\cal G} \rightarrow -2 \, \int d^3x \, \gamma^{1/2} \;
H \left( \chi | \chi_{(0)} \right ) \, ,
\end{equation}
which describes the integral over the original 3-geometry of the
the Hubble function
\begin{equation}
H \left( \chi | \chi_{(0)} \right ) = H_0 \, {\rm cotanh}
\left [ {3 H_0 \over 2} \left ( \chi - \chi_{(0)} \right ) \right ] \, .
\end{equation}
\endnumparts
The Hubble function describes gravity, dust and a cosmological constant
(the scalar field drops out),
and it was first computed by Salopek and Stewart (1992).

\section{Green function solution: gravity and scalar field with
exponential potential}

Consider a scalar field with exponential potential
\begin{equation}
V(\phi) = V_0 \, {\rm exp} \left ( - \sqrt { 2 \over p } \phi  \right ) \, .
\end{equation}
The reduced HJ equation (\ref{ecreduced}) becomes:
\begin{equation}
- {1 \over 12 } \, e^{-3 \alpha} \,
\left( {\delta {\cal G} \over \delta \alpha } \right )^2
+{1 \over 2 }  \, e^{-3 \alpha} \,
\left( {\delta {\cal G} \over \delta z} \right )^2
+{1 \over 2 }  \, e^{-3 \alpha} \,
\left( {\delta {\cal G} \over \delta \phi } \right )^2
+  e^{3 \alpha} \,
V_0 \, {\rm exp} \left ( - \sqrt { 2 \over p } \phi  \right ) = 0 \, .
\end{equation}
The Green function solution which depends on seven parameter fields
$[\gamma^{(0)}_{ab}(x), \phi_{(0)}(x)]$ is
\numparts
\begin{equation}
{\cal G} = -\sqrt{ 4 V_0 \over 3 \left ( 1 - 1/(3p) \right ) } \,
\int d^3x \,  \sqrt{f} \, ,
\end{equation}
where $f$ is an abbreviation for
\begin{equation}
f = \gamma \, e^{- \sqrt{2 \over p} \phi } +
\gamma_{(0)}\, e^{- \sqrt{2 \over p} \phi_{(0)} }
-2 \, \gamma^{1/2} \, \gamma_{(0)}^{1/2} \,
{\rm exp} \left[- {1\over \sqrt{2p} } \left( \phi + \phi_{(0)} \right)
\right ] \,
{\rm cosh} \, \theta \, ,
\end{equation}
and
\begin{equation}
\theta = \sqrt{ {3\over 2 } } \,
\sqrt{
\left [  \phi - \phi_{(0)}
- {1 \over \sqrt{18 p} }
\ln (  \gamma / \gamma_{(0)} )  \right ]^2
+ \left ( 1 - {1\over 3p} \right ) \, z^2
} \; .
\end{equation}
\endnumparts
The sign in front of the Green function is arbitrary and I have
chosen a negative sign to describe an expanding universe.

\subsection{$p \rightarrow \infty$}

As $p \rightarrow \infty$, the potential for the scalar field approaches
a constant, and one recovers the case of gravity, a massless scalar field
and a cosmological constant described by eq.(\ref{greenmc}).

\subsection{Special Case: $\gamma_{(0)} \rightarrow 0$ limit}

Note that as $\gamma_{(0)} \rightarrow 0$, the Green function approaches
\numparts
\begin{equation}
{\cal G} \rightarrow -2 \, \int d^3x \, \gamma^{1/2} \; H(\phi) \, ,
\end{equation}
where the Hubble function was computed by Salopek and Bond (1990):
\begin{equation}
H(\phi) = \sqrt{ V_0 \over 3 \left ( 1 - 1/(3p) \right ) } \;
{\rm exp}\left (  - { \phi \over  \sqrt{2p} }\right ) \, .
\end{equation}
\endnumparts

\section{Examples of Lagrange multiplier method: gravity, dust
and scalar field}

I will illustrate the Lagrange multiplier method by considering various cases
of gravity, dust, and a massless scalar field whose evolution is described by
the Green function eq.(\ref{greenff}).
The case with cosmological constant is also interesting but
the algebra is more complicated and I will devote section 7 to it.

The Lagrange multiplier method for solving the HJ equation had
been foreshadowed in an earlier paper (Salopek 1991).
I will show how to recover these earlier results from the
generalized formalism described in section 2.

In the present section, I will consider the special case
where the functionals ${\cal I}$ and ${\cal L}$ which specify the
initial setting in eq.(\ref{gil}) do not
contain any spatial gradients. Many examples are given.

\subsection{Initial setting contains no spatial gradients}

I will choose the function $f$ appearing in the Lagrange multiplier term
(\ref{Lagrange}) to be an arbitrary function of $\phi_{(0)}$ and
$\chi_{(0)}$:
\begin{equation}
f \equiv f \left ( \phi_{(0)}, \chi_{(0)} \right ) \, ,
\end{equation}
and I will choose the initial state
\begin{equation}
{\cal I} = \int d^3 x \, \gamma_{(0)}^{1/2} \;
g\left ( \phi_{(0)}(x),  \chi_{(0)}(x) \right ) \, ,
\end{equation}
to be an integral over the initial volume of some arbitrary function $g$
of $\phi_{(0)}$ and $\chi_{(0)}$. This system will be tractable because
no spatial gradients appear.

Minimization of the generating functional eq.(\ref{gil})
with respect to the initial 3-metric $\gamma^{(0)}_{ab}(x)$ implies
\numparts
\begin{eqnarray}
z = && 0 \, , \\
\gamma_{(0)}^{1/4} = &&\gamma^{1/4} \, {\rm cosh}
\left[ \sqrt{3 \over 8} \left( \phi - \phi_{(0)} \right ) \right ] \Bigg /
\left [ 1 - {3 \over 4}  \left( \chi - \chi_{(0)} \right )
( g + L f ) \right ]  \, .
\end{eqnarray}
\endnumparts
Substitution into eq.(\ref{gil}) yields the reduced generating
functional
\begin{eqnarray}
{\cal S} = && - {4 \over 3 } \int d^3 x \, \gamma^{1/2} \,
{ 1 \over \left ( \chi - \chi_{(0)} \right ) }  \nonumber \\
+  && {4 \over 3 } \int d^3 x \, \gamma^{1/2} \,
{ {\rm cosh}^2
\left[ \sqrt{3 \over 8} \left( \phi - \phi_{(0)} \right ) \right ]
\over \left( \chi - \chi_{(0)} \right )
\left[ 1 - {3\over 4} \left( \chi - \chi_{(0)} \right ) ( g + L f) \right ]
\label{greduced}
}
\end{eqnarray}
which is still subject to minimization with respect to
$\phi_{(0)}(x)$,  $\chi_{(0)}(x)$ and $L(x)$. Since the purpose of
$L$ is solely to impose the constraint $f=0$, the above generating
function is equivalent to:
\begin{eqnarray}
{\cal S} = && - {4 \over 3 } \int d^3 x \, \gamma^{1/2} \,
{ 1 \over \left ( \chi - \chi_{(0)} \right ) }  \nonumber \\
+  && {4 \over 3 } \int d^3 x \, \gamma^{1/2} \,
{ {\rm cosh}^2
\left[ \sqrt{3 \over 8} \left( \phi - \phi_{(0)} \right ) \right ]
\over \left( \chi - \chi_{(0)} \right )
\left[ 1 - {3\over 4} \left( \chi - \chi_{(0)} \right )  g \right ]
} \nonumber \\
+ && \int d^3 x \, \gamma^{1/2} \;
L(x) \,  f\left ( \phi_{(0)}(x),  \chi_{(0)}(x) \right ) \, , \nonumber \\
&& ({\rm minimized \; \; with \; \; respect \; \; to \; \; }
\phi_{(0)}(x),  \; \; \chi_{(0)}(x)\; \;  {\rm and} \; \; L(x) ) \, .
\label{greduced1}
\end{eqnarray}
Here, I have set $f= 0$ in eq.(\ref{greduced}), and then appended
a Lagrange multiplier term to it to recover the equivalent condition.

\subsubsection{Recovery of previous results}

The application of the Lagrange multiplier method
to gravitational systems was suggested in an earlier paper by Salopek
(1991). It will be shown how these special results may be derived
using the generalized methods presented in section 2.

For two or more scalar fields denoted by $\phi_a$, $a=1,..., n$, $a \ge 2$,
it was demonstrated in Salopek (1991)
that one could obtain a special
class of solutions to the long-wavelength problem by attempting
an Ansatz which was an integral over the volume element of the Hubble
function, $H(\phi_a)$:
\numparts
\begin{equation}
{\cal S}[\gamma_{ab}(x), \phi_a(x)] = - 2 \int d^3x \, \gamma^{1/2} \,
H\left[ \phi_a(x) \right ] \, ,
\label{iov}
\end{equation}
provided that the Hubble function $H(\phi_a)$ satisfies the
separated Hamilton-Jacobi (SHJE):
\begin{equation}
H^2 = {2\over 3} \sum_{a=1}^n
\left( {\partial H \over \partial \phi_a} \right )^2 + { V(\phi) \over 3 }
\, .
\label{SHJE}
\end{equation}
\endnumparts
If one is fortunate to find a solution,
\begin{equation}
\tilde H\left (\phi_a| \tilde \phi_a \right ) \, ,
\end{equation}
of the SHJE (\ref{SHJE}) which depends on $n$
parameters, $\tilde \phi_a$, one may construct another solution $H(\phi_a)$,
\numparts
\begin{equation}
H(\phi_a) = \tilde H\left (\phi_a| \tilde \phi_a \right ),
\end{equation}
by choosing the parameters
to minimize the n-parameter solution with respect $\tilde \phi_a$ assuming that
the parameters are subject to a constraint:
\begin{equation}
f(\tilde \phi_a) = 0 \, .
\end{equation}
\endnumparts
By introducing a Lagrange multiplier $\lambda$, one can write this new
solution as
\numparts
\begin{equation}
H = \tilde H\left (\phi_a| \tilde \phi_a \right ) +
\lambda \, f(\tilde \phi_a) \, ,
\label{ha}
\end{equation}
where $H$ is minimized with respect to $\lambda$ and $\tilde \phi_a$:
\begin{equation}
{\partial H \over \partial \lambda} = 0 \, , \quad
{\partial H \over \partial \tilde \phi_a} = 0 \, .
\label{hb}
\end{equation}
\endnumparts

Now in this paper, I have focussed on a single scalar field and
a single dust field, but all past experience has show that
a dust field can basically be treated as scalar field with a
different term appearing in the energy constraint. In this subsection,
for the purpose of
simplicity, I will set
$g(\phi_{(0)}, \chi_{(0)}) = 0$ in eq.(\ref{greduced1}), and then
obtain the following generating functional,
\begin{eqnarray}
{\cal S} = &&  {4 \over 3 } \int d^3 x \, \gamma^{1/2} \,
{ {\rm sinh}^2
\left[ \sqrt{3 \over 8} \left( \phi - \phi_{(0)} \right ) \right ]
\over \left( \chi - \chi_{(0)} \right )
} \nonumber \\
+ && \int d^3 x \, \gamma^{1/2} \;
L(x) \,  f\left ( \phi_{(0)}(x),  \chi_{(0)}(x) \right ) \, , \nonumber \\
&& ({\rm minimized \; \; with \; \; respect \; \; to \; \; }
\phi_{(0)}(x),  \; \; \chi_{(0)}(x)\; \;  {\rm and} \; \; L(x) ) \, .
\label{greduced2}
\end{eqnarray}
However, this generating functional is of the form of
eq.(\ref{iov}) and eq.(\ref{ha}) where
\numparts
\begin{equation}
H = \tilde H \left [ \phi, \chi| \phi_{(0)}, \chi_{(0)} \right ]
+ \lambda f\left ( \phi_{(0)}(x),  \chi_{(0)}(x) \right ) \, ,
\quad (\lambda = - L/2) \, ,
\end{equation}
is minimized with respect to $\lambda$, $\phi_{(0)}$ and $\chi_{(0)}$.
Here, $\tilde H$,
\begin{equation}
\tilde H \left [ \phi, \chi| \phi_{(0)}, \chi_{(0)} \right ] =
- {2 \over 3 } \, { {\rm sinh}^2
\left[ \sqrt{3 \over 8} \left( \phi - \phi_{(0)} \right ) \right ]
\over \left( \chi - \chi_{(0)} \right )
} \, ,
\end{equation}
satisfies the following SHJE:
\begin{equation}
\tilde H^2 = -{2\over 3} \, { \partial \tilde H \over \partial \chi} +
{2\over 3} \, \left ( { \partial \tilde H \over \partial \phi} \right )^2 \, .
\end{equation}
\endnumparts
Hence the generalized Lagrange multiplier method recovers the special
case considered in Salopek (1991) which was described for $n$
scalar fields in eqs.(\ref{ha}-b).

I will now consider two special cases for $f$ corresponding to initial
hypersurfaces where firstly the dust field is uniform and where
secondly the scalar field is uniform.

\subsection{Uniform dust field initially}

As was discussed before in section 2, if one takes
\numparts
\begin{equation}
f\left ( \phi_{(0)}(x),  \chi_{(0)}(x) \right ) = \chi_{(0)}
\end{equation}
then variation of the Lagrange multiplier implies that
\begin{equation}
\chi_{(0)} = 0 \, ,
\end{equation}
and that the initial hypersurface is one where the dust field
is uniformly zero. In this case, I will further assume that
$g$ is an arbitrary function of $\phi_{(0)}$:
\begin{equation}
g \equiv g ( \phi_{(0)} ) \, .
\end{equation}
\endnumparts
One finds that the generating functional eq.(\ref{greduced1}) is:
\numparts
\begin{eqnarray}
{\cal S}[\gamma_{ab}(x), \phi(x)] = &&
{4 \over 3 } \int d^3 x \, \gamma^{1/2} \, {1 \over \chi} \,
\left \{
{
{\rm cosh}^2
\left[ \sqrt{3 \over 8} \left( \phi - \phi_{(0)} \right ) \right ]
\over \left( 1 - {3\over4} \chi g \right ) }
- 1
\right \} \; , \\
{\rm tanh}\left[ \sqrt{3 \over 8} \left( \phi - \phi_{(0)} \right ) \right ]
&&=
{
\sqrt{3 \over 8} \, \chi \, {\partial g \over \partial \phi_{(0)} }
\over \left( 1 - {3\over 4} \chi g \right )
} \; .
\end{eqnarray}
\endnumparts
The second equation is the minimization condition of the first with respect to
$\phi_{(0)}$.

For various choices of $g$, I have computed the resulting generating
functional.

For $g$ a constant:
\numparts
\begin{equation}
g = C \quad {\rm (a \; \; constant)} \, ,
\end{equation}
\begin{equation}
{\cal S}[\gamma_{ab}(x), \chi(x) ] = - { 4\over 3 } \,
\int d^3x \, \gamma^{1/2} \, {1 \over \left [ \chi - 4/(3C) \right ] } \, .
\end{equation}
\endnumparts

For $g$ proportional to an exponential:
\numparts
\begin{equation}
g = C \, {\exp} \left( \sqrt{3\over 2} \phi_{(0)} \right ) \, ,
\end{equation}
\begin{equation}
{\cal S}[\gamma_{ab}(x), \chi(x) ] = C \,
\int d^3 x \gamma^{1/2} \, {\exp} \left( \sqrt{3\over 2} \phi \right )
\, . \label{gexp}
\end{equation}
\endnumparts

For $g$ proportional to ${\rm cosh}^2(\sqrt{3/8} \, \phi_{(0)})$:
\numparts
\begin{equation}
g = C \, {\rm cosh}^2 \left( \sqrt{3\over 8} \phi_{(0)} \right ) \, ,
\end{equation}
\begin{equation}
{\cal S}[\gamma_{ab}(x), \chi(x) ] = - {4\over 3 } \,
\int d^3 x \gamma^{1/2} \,
{ {\rm cosh}^2 \left( \sqrt{3\over 8} \phi \right ) \over
\left [ \chi - 4/(3C) \right ] } \, .
\end{equation}
\endnumparts

Thus the Green function method leads to many solutions of strongly coupled
gravitational systems.

\subsection{Uniform scalar field initially}

Setting $f = \phi_{(0)}$ implies that the scalar field is uniformly
zero on the initial hypersurface. In this case, I now assume
that g is an arbitrary function of $\chi_{(0)}$:
\begin{equation}
g \equiv g \left ( \chi_{(0)} \right ) \, .
\end{equation}
In this case, the generating functional eq.(\ref{greduced1}) becomes:
\numparts
\begin{eqnarray}
{\cal S} = && - {4 \over 3 } \int d^3 x \, \gamma^{1/2} \,
{ 1 \over \left ( \chi - \chi_{(0)} \right ) }  \nonumber \\
+  && {4 \over 3 } \int d^3 x \, \gamma^{1/2} \,
{ {\rm cosh}^2
\left( \sqrt{3 \over 8} \phi \right )
\over \left( \chi - \chi_{(0)} \right )
\left[ 1 - {3\over 4} \left( \chi - \chi_{(0)} \right )  \, g  \right ]
} \; ,
\label{usinit}
\\
{\rm tanh}^2 && \left( \sqrt{3 \over 8} \phi \right) =
{\left( \chi - \chi_{(0)} \right )^2 \over
\left[ 1 - {3 \over 4}   \left( \chi - \chi_{(0)} \right ) g \right ]^2 } \,
\left(
{9 \over 16} g^2 - {3\over 4} {\partial g \over \partial \chi_{(0)}}
\right ) \, .
\label{minchi0}
\end{eqnarray}
\endnumparts
Equation (\ref{minchi0}) is the minimization condition of eq.(\ref{usinit})
with respect to $\chi_{(0)}$.

For various choices of $g$, I have computed the resulting generating
functional.

For $g$ a constant:
\numparts
\begin{equation}
g = C \quad {\rm (a \; constant)} \, ,
\end{equation}
\begin{equation}
{\cal S}[\gamma_{ab}(x), \chi(x) ] = C \,
\int d^3 x \gamma^{1/2} \, {\exp} \left( \sqrt{3\over 2} \phi \right )
\, .
\end{equation}
which is the same result found in eq.(\ref{gexp}) using a different route.
\endnumparts

For $g = -4 /(3 \chi_{(0)})$, one obtains
\numparts
\begin{equation}
g = - {4 \over 3 } \, {1 \over \chi_{(0)} } \, , \label{g2}
\end{equation}
\begin{equation}
{\cal S}= - {4 \over 3} \, \int d^3 x \, \gamma^{1/2} \, {1 \over \chi}  \, .
\end{equation}
\endnumparts

For $g = C/ \chi_{(0)}$,
\numparts
\begin{equation}
g =  {C \over \chi_{(0)} } \, ,
\end{equation}
\begin{equation}
{\cal S}=  {4 \over 3} \, \int d^3 x \, \gamma^{1/2} \,
{ {\rm sinh}^2\left( \sqrt{3\over 8} \phi + \beta \right ) \over \chi}  \, .
\end{equation}
where
\begin{equation}
{\rm sinh}^2 \beta = {3 C \over 4 } \, .
\end{equation}
\endnumparts
What is rather unusual is that for
\numparts
\begin{equation}
C=  - {4\over 3} \,
\end{equation}
one finds $\beta = i \pi/2$, and the resulting generating functional is
\begin{equation}
{\cal S}=  - {4 \over 3} \, \int d^3 x \, \gamma^{1/2} \,
{ {\rm cosh}^2\left( \sqrt{3\over 8} \phi \right ) \over \chi}  \, ,
\end{equation}
\endnumparts
which is certainly a solution to the energy and momentum constraints
but it is not
the solution given in eq.(\ref{g2},b)! However, note that when
$\phi=0$, they do agree ---- they share the same initial functional,
${\cal I}$. These two  results reflect the multivalued solutions of
the minimization equation (\ref{minchi0}).

\section{Lagrange multiplier method: gravity, dust, massless scalar field
with cosmological constant}

I will repeat the analysis of section 6 but I will now add a cosmological
constant to the system of gravity, dust and a massless scalar field.
The Green function ${\cal G}$ was given in eq.(\ref{gdmsc}).

When the functionals, ${\cal I}$ and ${\cal L}$,
\numparts
\begin{equation}
{\cal I} = \int d^3 x \, \gamma_{(0)}^{1/2} \;
g\left ( \phi_{(0)}(x),  \chi_{(0)}(x) \right ) \, ,
\end{equation}
\begin{equation}
{\cal L} = \int d^3 x \, \gamma_{(0)}^{1/2} \; L(x) \,
f\left ( \phi_{(0)}(x),  \chi_{(0)}(x) \right ) \, ,
\end{equation}
\endnumparts
which determine the initial setting do not contain spatial gradients,
then the minimization of ${\cal S}$, eq.(\ref{gil}),
with respect to the initial 3-metric
is straightforward:
\numparts
\begin{equation}
z = 0 \, ,
\end{equation}
\begin{equation}
\left( {\gamma_{(0)} \over \gamma } \right )^{1/4} =
{ {\rm cosh} \left[ \sqrt{3 \over 8} \left( \phi - \phi_{(0)} \right ) \right ]
\over
{\rm cosh}
\left[ {3 H_0 \over 2 } \, \left( \chi - \chi_{(0)} \right ) \right ]
- { (g + Lf) \over  2 \, H_0 }\,
{\rm sinh}
\left[ {3 H_0 \over 2 } \left( \chi - \chi_{(0)} \right ) \right ] } \, ,
\end{equation}
\endnumparts
giving
\begin{eqnarray}
&&{ {\cal S} \over (2 H_0) }  = - \int d^3x \, \gamma^{1/2} \,
{\rm cotanh}
\left[ {3 H_0 \over 2 } \, \left( \chi - \chi_{(0)} \right ) \right ] +
\nonumber \\
&&
\int d^3x \, \gamma^{1/2} \, {
{\rm cosh}^2 \left[ \sqrt{3 \over 8} \left( \phi - \phi_{(0)} \right ) \right ]
\over
{\rm sinh}
\left[ {3 H_0 \over 2 } \left( \chi - \chi_{(0)} \right ) \right ]
} \;
{1
\over
{\rm cosh}
\left[ {3 H_0 \over 2 } \left( \chi - \chi_{(0)} \right ) \right ]
- { g\over 2 H_0 } \;
{\rm sinh}
\left[ {3 H_0 \over 2 } \left( \chi - \chi_{(0)} \right ) \right ]
}  \nonumber \\
&& + {1 \over (2 H_0) } \, \int d^3 x \, \gamma^{1/2} \, L \, f \, .
\end{eqnarray}
In the above equation, I followed the analysis of section 6
by setting $f = 0$ in the generating functional and then
recovering this same condition  by adding a Lagrange multiplier term which is
a linear functional in $L(x)$.

\subsection{Uniform dust field initially}

I will assume that initial state is given on an initial hypersurface
where the dust field is zero which is characterized by a vanishing
value of $\chi_{(0)}(x)$:
\begin{equation}
f = \chi_{(0)}(x) = 0 \, .
\end{equation}
The Lagrange multiplier term will thus be removed.
Moreover, I will assume that $g$, which characterizes the initial state, is
a function soley of the initial scalar field $\phi_{(0)}$:
\begin{equation}
g \equiv g \left(  \phi_{(0)} \right ) \, .
\end{equation}
After minimizing with respect to $ \phi_{(0)}$, the generating functional is:
\numparts
\begin{eqnarray}
{\cal S}&&[\gamma_{ab}(x), \phi(x), \chi(x) ] / (2 H_0)=
- \int d^3x \, \gamma^{1/2} \, {\rm cotanh} \theta \nonumber + \\
&&
\int d^3x \, \gamma^{1/2} \, {1\over {\rm sinh}^2 \theta } \;
{
{\rm cotanh} \theta -h\left( \phi_{(0)} \right )
\over
\left[ {\rm cotanh} \theta - h\left( \phi_{(0)} \right ) \right ]^2 -
{2\over 3} \left( {dh \over d \phi_{(0)} } \right )^2
} \; ,
\end{eqnarray}
where $\theta$ and $h$ are dimensionless representations of $\chi$ and $g$,
\begin{equation}
\theta \equiv { 3 H_0 \chi \over 2 } \, , \quad h \equiv
g\left( \phi_{(0)} \right ) / (2 H_0 )  \, . \label{defgtheta}
\end{equation}
The parameter field $\phi_{(0)}(x)$ is determined implicitly through the
algebraic equation,
\begin{equation}
{\rm tanh}
\left [ \sqrt{3 \over 8} \left( \phi - \phi_{(0)}(x) \right ) \right ] =
{
\sqrt{2 \over 3} \,  {dh \over d \phi_{(0)} }
\over
{\rm cotanh} \theta - h \left( \phi_{(0)} \right )
} \, . \label{phi0implicit}
\end{equation}
\endnumparts

\subsubsection{Exact solution}

One obtains a non-trivial exact solution if one assumes that $g$ has the
following form:
\numparts
\begin{equation}
{ g \left( \phi_{(0)} \right )  \over 2 H_0} \equiv  h = C \,
{\rm exp} \left( - \sqrt{3\over2} \, \phi_{(0)} \right ) \,  +
D + E \, {\rm exp} \left(  \sqrt{3\over2} \, \phi_{(0)} \right ) \, .
\end{equation}
One can then solve $\phi_{(0)}$ explicitly using eq.(\ref{phi0implicit}), and
the generating functional is
\begin{equation}
{\cal S}[\gamma_{ab}(x), \phi(x), \chi(x) ] =
- 2 \, \int d^3x \, \gamma^{1/2} \, H(\phi, \chi) \, ,
\end{equation}
where the Hubble function $H(\phi, \chi)$ is given by
\begin{eqnarray}
&& H(\phi, \chi) / H_0=
\, {\rm cotanh} \theta \nonumber  \\
&& - {1\over  {\rm sinh}^2 \theta } \;
{
\left [ E b + C/ b - D + {\rm cotanh} \theta \right ]
\over \left [ ({\rm cotanh}\theta - D)^2 - 4 \, C E \right ]
} \;  ,
\end{eqnarray}
and $b$ denotes
\begin{equation}
b =  {\rm exp} \left( \sqrt{3\over2} \, \phi \right ) \, .
\end{equation}
\endnumparts
As a check of this method, one may verify that $H$ is a solution of the
SHJE,
\begin{equation}
H^2 = -{2\over 3} \, { \partial H \over \partial \chi} +
{2\over 3} \, \left ( { \partial H \over \partial \phi} \right )^2 +
{ V_0 \over 3 } \, .
\end{equation}

Some special cases illuminate the solution.

If $D = C = E = 0$, the Hubble function is
\begin{equation}
H = H_0 \, {\rm tanh} \left( {3 H_0 \chi \over 2} \right ) \, ,
\quad (D= C = E = 0) \, .
\end{equation}

If $D = 0$, $C = E = - {1\over 2}$, the Hubble function is
\begin{equation}
H = H_0 \, {\rm cosh} \left ( \sqrt{3\over 2} \, \phi \right ) \, ,
\quad (D=0, \; C = E = - {1\over 2} ) .
\end{equation}

If $D=0$ and $C= E$, the Hubble function is
\begin{equation}
{H \over H_0 }=
{
\left( 1 - 4 E^2 \right ) {\rm cosh}\theta \, {\rm sinh} \theta
- 2 E \, {\rm cosh}\left( \sqrt{3\over 2} \phi \right )
\over
{\rm cosh}^2 \theta - 4 E^2 \, {\rm sinh}^2 \theta
} \, ,
\quad (D=0, \; C = E ) \, .
\end{equation}
\noindent

\section{Advanced example:
initial hypersurface condition containing spatial gradients}

I will now consider an advanced example where the Lagrange multiplier term
contains spatial gradients. This example is particularly important because
it illustrates how the energy constraint applies to the initial fields.

The system under examination will contain only
gravity and dust, and the Green function is given by eq.(\ref{greenf}-c).
The function $f$ which defines the initial hypersurface will be chosen to be,
\numparts
\begin{equation}
f = \chi_{(0)} -  B R_{(0)} \, ,
\label{m_advanced}
\end{equation}
where $B$ is a constant and $R_{(0)}$ is the Ricci scalar associated with
the initial 3-metric. The functional ${\cal I}$ is taken to be
the simplest non-trivial example, namely the volume of the initial 3-geometry:
\begin{equation}
{\cal I} = A \, \int d^3 x \, \gamma_{(0)}^{1/2} \, .
\end{equation}
\endnumparts
The full generating functional is the sum of ${\cal G}$, ${\cal I}$ and
${\cal L}$:
\begin{eqnarray}
{\cal S}&& [ \gamma_{ab}(x), \chi(x) ]
=  \nonumber \\ && {4 \over 3 } \int d^3 x \;
{1 \over \left ( \chi(x) - \chi_{(0)}(x) \right ) } \;
 \left [ 2 \gamma^{1/4} \; \gamma_{(0)}^{1/4} \;
{\rm cosh} ( \sqrt{3 \over 8} z )
- \gamma^{1/2} - \gamma_{(0)}^{1/2} \right ] \, , \nonumber \\
+  && A \, \int d^3 x \, \gamma_{(0)}^{1/2} +
\int d^3x \, \gamma_{(0)}^{1/2} \,
L(x) \, \left [ \chi_{(0)}(x) - B R_{(0)}(x) \right ]  \nonumber \\
&& \left( {\rm minimized \; \; with \; \; respect \; \; to \; \;}
\gamma^{(0)}_{ab}(x), \; \; \chi_{(0)}(x) \; \; {\rm and} \; \; L(x) \; \;
\right )\, .
\label{fulls}
\end{eqnarray}
This example will not be exactly solvable unless $B$ vanishes.
One obtains approximate results by expanding in powers of $B$.

\subsection{Minimization conditions}

Variation with respect to $L(x)$ and $\chi_{(0)}$ lead immediately to:
\numparts
\begin{equation}
0 = \chi_{(0)} - B R_{(0)} \, ,
\label{var1}
\end{equation}
\begin{equation}
L = - {4 \over 3 }
{
\left [ 2 \left( \gamma / \gamma_{(0)}  \right )^{1/4} \,
	{\rm cosh}\left ( \sqrt{3\over 8} z \right ) -
	\left( \gamma / \gamma_{(0)} \right )^{1/2} - 1 \right ] \over
	\left( \chi - \chi_{(0)} \right )^2
} \, .
\label{var2}
\end{equation}
After minimizing eq.(\ref{fulls}) with respect to $\gamma^{(0)}_{ab}$, the
trace and traceless parts are (the trace is defined using the initial
3-metric):
\begin{equation}
0 = 2 { \left ( \left( \gamma / \gamma_{(0)}  \right )^{1/4} \,
   {\rm cosh} \left ( \sqrt{3\over 8} z \right ) - 1 \right ) \over
	\chi - \chi_{(0)} }
+ { 3A \over 2 } + B \left ( R_{(0)} \, L + 2 L^{;c}{}_{;c} \right ) \, ,
\label{var3}
\end{equation}
\begin{eqnarray}
0 = && 6^{-1/2} \,  { \left( \gamma / \gamma_{(0)}  \right )^{1/4} \over
		\left( \chi - \chi_{(0)} \right ) } \;
	{ {\rm sinh}\left ( \sqrt{3\over 8} z \right ) \over
	     z } \;
	\left [ \ln \left( [h^{-1}] \, [h_{(0)}] \right ) \right ]^a{}_b
	\nonumber \\
   && B \left ( \overline R_{(0)}^a{}_b - L^{;a}{}_{;b} +
	{1\over 3} \, \delta^{a}_{b} \, L^{;c}{}_{;c}  \right ) \, .
\label{var4}
\end{eqnarray}
\endnumparts

\subsection{Initial energy constraint}

After one minimizes ${\cal S}$, the initial fields, $\gamma^{(0)}_{ab}(x)$,
$\chi_{(0)}(x)$ and $L(x)$, are related in a subtle way.
By construction,  the Green function ${\cal G}$ satisfies the energy
constraint:
\begin{equation}
0 = { \delta {\cal G} \over \delta \chi(x) } +
\gamma^{-1/2} \, \left ( 2 \gamma_{ac} \, \gamma_{bd} -
\gamma_{ab} \, \gamma_{cd} \right ) \,
{ \delta {\cal G} \over \delta \gamma_{ab}(x) } \,
{ \delta {\cal G} \over \delta \gamma_{cd}(x) }  \, .
\label{ec_original}
\end{equation}
We interpret this equation as stating that the original fields satisfy
the energy constraint.
However, because of a symmetry transformation that relates
the original fields and the initial fields,
\begin{equation}
\Big ( \gamma_{ab}(x), \chi(x) \Big ) \leftarrow \rightarrow
\Big ( \gamma^{(0)}_{ab}(x), - \chi_{(0)}(x) \Big ) \, ,
\end{equation}
the initial fields, $\gamma^{(0)}_{ab}(x)$ and $\chi_{(0)}(x)$,
also satisfy the energy constraint:
\begin{equation}
0 = -{ \delta {\cal G} \over \delta \chi_{(0)}(x) } +
\gamma_{(0)}^{-1/2} \, \left ( 2 \gamma^{(0)}_{ac} \, \gamma^{(0)}_{bd} -
\gamma^{(0)}_{ab} \, \gamma^{(0)}_{cd} \right ) \,
{ \delta {\cal G} \over \delta \gamma^{(0)}_{ab}(x) } \,
{ \delta {\cal G} \over \delta \gamma^{(0)}_{cd}(x) }  \, .
\label{ec_initial}
\end{equation}
By virtue of the minimization conditions (see,{\it e.g.}, eq.(\ref{la}-d)),
one may thus replace functional derivatives of ${\cal G}$ with functional
derivatives of $({\cal I + L})$,
\numparts
\begin{eqnarray}
-{ \delta {\cal G} \over \delta \chi_{(0)}(x) } &&=
{ \delta ({\cal I + L}) \over \delta \chi_{(0)}(x) } \, , \\
-{ \delta {\cal G} \over \delta \gamma^{(0)}_{ab}(x) } &&=
{ \delta ({\cal I + L}) \over \delta \gamma^{(0)}_{ab}(x) } \, ,
\end{eqnarray}
in eq.(\ref{ec_initial}) which leads to the {\it initial energy constraint},
\endnumparts
\begin{eqnarray}
0 &&= { \delta ({\cal I + L}) \over \delta \chi_{(0)}(x) } +
\gamma_{(0)}^{-1/2} \, \left ( 2 \gamma^{(0)}_{ac} \, \gamma^{(0)}_{bd} -
\gamma^{(0)}_{ab} \, \gamma^{(0)}_{cd} \right ) \,
{ \delta ({\cal I+ L}) \over \delta \gamma^{(0)}_{ab}(x) } \,
{ \delta ({\cal I + L}) \over \delta \gamma^{(0)}_{cd}(x) }
\label{iec1} \\
&&\quad ({\rm initial \; \; energy \; \; constraint})  \, , \nonumber
\end{eqnarray}
which is independent of the original fields, $\gamma_{ab}(x)$ and $\chi(x)$.
Explicit evaluation of the functional derivatives in eq.(\ref{iec1}) yields
a relationship amongst the Lagrange multiplier and the initial
fields:
\numparts
\begin{equation}
0 = L + 2 B^2 \overline m^{ab} \, \overline m_{ab}
- {1 \over 3 } \left ( B \, m + {3\over 2} A \right )^2 \, ,
\quad ({\rm initial \;  energy \; constraint})
\label{iec2} \\
\end{equation}
where the tensor $m^{ab}$ is given by
\begin{equation}
m^{ab} = R_{(0)}^{ab} \, L  + \gamma_{(0)}^{ab} \, L^{;c}{}_{;c}
- L^{;ab} \;  \, ,
\end{equation}
and $\overline m^{ab}$ denotes its traceless component:
\begin{equation}
\overline m^{ab} = m^{ab} - {1\over 3 } \gamma_{(0)}^{ab} \, m \, ,
\quad {\rm with} \quad m = \gamma^{(0)}_{ab} \, m^{ab} \, .
\end{equation}
\endnumparts
In the above, a semicolon $(;)$ denotes a derivative with respect to the
initial
3-metric and all indices are raised and lowered using the initial 3-metric.
In principle, one may solve for the Lagrange multiplier in terms of the
initial 3-metric, although this is difficult to achieve in practice since
this is a fourth order, nonlinear partial differential equation.

\subsection{Classical evolution}

Classical evolution is found by solving eqs.(\ref{var1}-d) for
$\gamma_{ab}(x)$ in terms of the proper time $\chi(x)$, the initial fields,
$\gamma^{(0)}(x)$ and $\chi_{(0)}(x)$,
and the Lagrange multiplier, $L(x)$:
\numparts
\begin{equation}
\left( {\gamma \over \gamma_{(0)} }\right )^{1/2} =
\left [ 1 - { \left ( \chi - \chi_{(0)} \right ) \over
              \left ( \chi_{(1)} - \chi_{(0)} \right ) } \right ] \;
\left [ 1 - { \left ( \chi - \chi_{(0)} \right ) \over
              \left ( \chi_{(2)} - \chi_{(0)} \right ) } \right ] \, ,
\label{class_a}
\end{equation}
\begin{equation}
[h] = [h^{(0)}] \, {\rm exp} \left ( 2 z { [\overline m] [\gamma_{(0)}]
\over  \sqrt{ \overline m^{ab} \, \overline m_{ab} } } \right ) \, ,
\end{equation}
\begin{equation}
z = \sqrt{2 \over 3} \, \ln
\left [ {  1 - { \left ( \chi - \chi_{(0)} \right ) \over
              \left ( \chi_{(2)} - \chi_{(0)} \right ) } \over
1 - { \left ( \chi - \chi_{(0)} \right ) \over
              \left ( \chi_{(1)} - \chi_{(0)} \right ) } }
\right ] \, .
\end{equation}
$\left( \gamma / \gamma_{(0)} \right )^{1/2}$ is a quadratic function
of $\left ( \chi - \chi_{(0)} \right )$. The terms,
$\left ( \chi_{(1)} - \chi_{(0)} \right )$ and
$\left ( \chi_{(2)} - \chi_{(0)} \right )$, denote the two roots of
$\left( \gamma/ \gamma_{(0)} \right )^{1/2}$:
\begin{eqnarray}
\left ( \chi_{(1)} - \chi_{(0)} \right ) &&=
{A \over  L} +
{2 B \over 3} \, \left( R_{(0)} + {2 \over L} \, L^{;c}{}_{;c} \right )
- \sqrt{8\over 3} { B \over L } \,
\sqrt{ \overline m^{ab} \, \overline m_{ab} }  \, ,
\label{root1} \\
\left ( \chi_{(2)} - \chi_{(0)} \right ) &&=
{A \over  L} +
{2 B \over 3} \, \left( R_{(0)} + {2 \over L} \, L^{;c}{}_{;c} \right )
+\sqrt{8\over 3} { B \over L } \,
\sqrt{ \overline m^{ab} \, \overline m_{ab} }
 \, .
\label{root2}
\end{eqnarray}
\endnumparts
The tensor $m^{ab}$ was given earlier but I will write it down here just
to have all of the results in one place:
\begin{equation}
m^{ab} = R_{(0)}^{ab} \, L  + \gamma_{(0)}^{ab} \; L^{;c}{}_{;c}
- L^{;ab}  \, .
\end{equation}
$\chi_{(0)}$ is given by the initial hypersurface condition:
\begin{equation}
0 = \chi_{(0)} - B R_{(0)} \, ,
\label{ihsc}
\end{equation}
and the Lagrange multiplier may be solved in terms of the initial 3-metric
from the initial energy constraint:
\begin{equation}
0 = L + 2 B^2 \overline m^{ab} \, \overline m_{ab}
- {1 \over 3 } \left ( B \, m + {3\over 2} A \right )^2 \, , \\
\end{equation}

One may solve for $L$, and subsequently all other quantities, by expanding in
powers of $B$. To first order in $B$, one finds:
\numparts
\begin{equation}
L = {3 A^2 \over 4} \left( 1 + B A R_{(0)} \right ) \, ,
\end{equation}
\begin{equation}
\chi_{(0)} = B R_{(0)} \, ,
\end{equation}
\begin{equation}
m^{ab} = \left( {3 A^2 \over 4} \right ) \,
\left [  R_{(0)}^{ab} + B A \,
\left ( R_{(0)} \,   R_{(0)}^{ab} - R_{(0)}^{;ab} +
\gamma_{(0)}^{ab} \, \tilde D^2 R_{(0)} \right ) \right ] \, ,
\end{equation}
\begin{equation}
m = \left( {3 A^2 \over 4} \right ) \,
\left [ R_{(0)} + B A \,
\left ( R_{(0)}^2 + 2 \tilde D^2 R_{(0)} \right ) \right ]
\, ,
\end{equation}
\begin{equation}
\chi_{(1)} -  \chi_{(0)} = \left( {4 \over 3 A} \right )
\left( 1 - {B A \over 2} \, R_{(0)} \right )
- \sqrt{8\over 3} \, B \,
\left ( \overline R_{(0)}^{ab} \overline R^{(0)}_{ab} \right )^{1/2} \, ,
\end{equation}
\begin{equation}
\chi_{(2)} -  \chi_{(0)} = \left( {4 \over 3 A} \right )
\left( 1 - {B A \over 2} \, R_{(0)} \right )
+ \sqrt{8\over 3} \, B \,
\left ( \overline R_{(0)}^{ab} \overline R^{(0)}_{ab} \right )^{1/2} \, ,
\end{equation}
\begin{equation}
z = {3 B A^2 \over 2 } \,
{ \chi \over \left( 1 - {3 A \chi \over 4} \right ) } \,
\left ( \overline R_{(0)}^{ab} \overline R^{(0)}_{ab} \right )^{1/2} \, ,
\end{equation}
\begin{equation}
\gamma_{ab} = \left( 1 - { 3 A \chi \over 4 } \right )^{4/3} \,
\left [ \gamma_{ab}^{(0)} +
{ A B \over \left( 1 - { 3 A \chi \over 4 } \right ) } \,
\left ( 3 A \chi \, R^{(0)}_{ab} + ( 1 - { 3 A \chi \over 2 } ) \,
R_{(0)} \, \gamma^{(0)}_{ab} \right ) \right ] \, .
\label{first.few.terms}
\end{equation}
\endnumparts
In the above, the laplacian of $R_{(0)}$ is denoted by
\begin{equation}
\tilde D^2 R_{(0)} \equiv  \left( R_{(0)} \right )^{;c}{}_{;c} \, .
\end{equation}

\subsection{Determination of the classical 4-metric}

In solving the classical Einstein equations, one ordinarily computes the
4-metric describing time and space. However, in the HJ formalism, one's
attention is primarily focussed on the 3-metric which describes the
spatial geometry. How does one recover the 4-metric? In general,
given the generating functional ${\cal S}[\gamma_{ab}(x), \chi(x)]$,
the 4-metric is computed by making an arbitrary choice for the time
parameter, and then integrating the definition of the momenta,
\numparts
\begin{equation}
\left(\dot\gamma_{ij}-N_{i|j}-N_{j|i}\right)/N =2 \kappa \, \gamma^{-1/2}
\left(2\gamma_{ik}\gamma_{jl}-\gamma_{ij}\gamma_{kl}\right)
{ \delta {\cal S} \over \delta \gamma_{kl} } \, ,
\label{metric.evol}
\end{equation}
\begin{equation}
\left(\dot\chi - N^i \chi_{,i}\right)/N= \kappa \, ,
\label{dust.evol}
\end{equation}
\endnumparts
which is valid in the strongly coupled limit.

For the problems considered in this section, a natural choice for the
time hypersurface is one where the dust field $\chi$ is uniform,
\numparts
\begin{equation}
t = \chi/\kappa
\end{equation}
which describes a comoving slice. If one assumes that
the shift $N_i$ vanishes, then eq.(\ref{dust.evol}) implies that the
lapse $N$ is equal to one,
\begin{equation}
N =1,
\end{equation}
\endnumparts
which describes a synchronous gauge. The line-element describing the
4-geometry is then
\begin{equation}
ds^2 = - {d \chi^2 \over \kappa^2} + \gamma_{ab} \, dx^a \, dx^b \, ,
\end{equation}
where the classical evolution of the 3-metric $\gamma_{ab}$ was computed
in the previous section using the HJ minimization prescription;
see eq.(\ref{first.few.terms}) for the first few terms.
At each spatial point, the 4-geometry is locally Kasner
(see, for example, Salopek 1998).

\subsection{Computation of generating functional}

Using the exact expressions for classical evolution, eqs.(\ref{class_a}-e),
one may express
the generating functional ${\cal S}$, eq.(\ref{fulls}), in an
elegant form which depends on the dust field $\chi(x)$, the initial
fields and the Lagrange multiplier:
\begin{equation}
{\cal S} = - \int d^3x \, \gamma_{(0)}^{1/2} \, L \,
\left ( \chi - \chi_{(0)} \right ) + {\cal I} + {\cal L}\, .
\label{elegant}
\end{equation}
Here I have eliminated all reference to the original 3-metric
$\gamma_{ab}(x)$.
As was found in Salopek (1998), the generating functional is basically linear
in $\chi(x)$. If $L(x)$ is positive, the integrand of ${\cal S}$
decreases in $\chi(x)$, whereas the opposite is true if $L(x)$ is negative.

Ultimately, one wishes to compute ${\cal S}$ soley as a functional
of the original fields, $\gamma_{ab}(x)$ and $\chi(x)$. Since this
cannot be done exactly for the problem at hand, I will be content to
approximate it using
a Taylor series in $B$. First note that the initial 3 metric,
$\gamma^{(0)}_{ab}(x)$ may be expressed as a function of
$\gamma_{ab}(x)$ and $\chi(x)$:
\numparts
\begin{equation}
\gamma^{(0)}_{ab} = k_{ab} -
{A B \over \left( 1 - {3 A \chi \over 4 } \right ) } \,
\left [ 3 A \, \chi \, R^k_{ab} + \left ( 1 - {3 A \chi \over 2 } \right ) \,
R^k \, k_{ab} \right ] \, ,
\end{equation}
which is accurate to first order in $B$. Here the tensor $k_{ab}$
is conformally related to the original 3-metric, $\gamma_{ab}$,
\begin{equation}
k_{ab} = \left( 1 - {3 A \chi \over 4 } \right )^{-4/3} \, \gamma_{ab} \, ,
\end{equation}
\endnumparts
and $R^k$, $R^k_{ab}$ denote the corresponding Ricci scalar and tensor,
respectively. After imposing the initial hypersurface condition,
eq.(\ref{ihsc}), one may safely drop the Lagrange multiplier term
${\cal L}$ in
eq.(\ref{elegant}) and the generating functional becomes,
\begin{equation}
{\cal S}[\gamma_{ab}(x), \chi(x) ] = A \int d^3x \, \gamma^{1/2} \,
{1 \over  \left( 1 - {3 A \chi \over 4 } \right ) } -
{ 3 A^2 B \over 4 } \int d^3 x \; k^{1/2} \, R^k \, ,
\end{equation}
which is accurate to first order in $B$. Conceptually, there is
no problem in extending this expression to higher order in $B$.

\section{Summary and conclusions}

A strong coupling expansion appears in many investigations of gravitational
systems including long-wavelength cosmological fluctuations,
gravitational collapse and string theory formulations of cosmology.
In the present paper,
powerful Hamilton-Jacobi methods have been further developed to allow for
a very general solution of strongly coupled gravitational systems.

In section 2, one generalizes the {\it semiclassical Green function method}
of solving the constraint equations to encompass the specification of an
arbitrary initial hypersurface.
One assumes an {\it Ansatz} for the generating functional ${\cal S}$,
eq.(\ref{gil}),  which is the sum of three terms:
\numparts
\begin{equation}
{\cal S} = {\cal G} + {\cal I} + {\cal L} \, .
\label{summary}
\end{equation}
The Lagrange multiplier term,
\begin{equation}
{\cal L} = \int d^3 x \, \gamma_{(0)}^{1/2} \; L(x) \,
f\left [ \gamma^{(0)}_{ab}(x), \phi_{(0)}(x),  \chi_{(0)}(x) \right ] \, ,
\label{summary1}
\end{equation}
specifies the initial hypersurface,
\begin{equation}
f= 0 \, .
\label{summary3}
\end{equation}
\endnumparts
${\cal I}$ is the initial state.
The Green function ${\cal G}$ determines how the systems evolves from
the initial setting.
The {\it Ansatz} (\ref{summary}) is justified mathematically by computing
its functional
derivatives. It hence satisfies the energy constraint.  Because
gauge-invariance is maintained at each step, the {\it Ansatz} satisfies
the momentum constraint. Classical evolution follows from minimization
of the generating functional (\ref{summary}) with respect to the
initial fields and the Lagrange multiplier $L$.

To illustrate the generalized method, I constructed in sections 4 and 5
Green function solutions
for (1) gravity, dust, a massless scalar field and a cosmological constant
and (2) gravity interacting with a scalar field with exponential potential.
One verifies these solutions after deriving a {\it reduced energy
constraint} in section 3.

If the functionals, ${\cal I}$ and ${\cal L}$, which define the
initial setting do not contain spatial gradients of the initial fields,
then the program is straightforward to implement, and one recovers
a primitive form of the minimization principle that had been advanced
in an earlier paper (Salopek 1991). In fact, in sections 6 and 7,
exact solutions were derived that had not been previously known.

However, in general the initial
hypersurface condition (\ref{summary3})
contains spatial gradients, and one must typically resort to an approximation
method as was illustrated in section 8 for the case of gravity and dust.
There it was shown explicitly how to deal with the initial energy constraint.

The {\it semiclassical Green function method}
has reached a high level of generality which should be sufficiently
powerful to treat most strongly coupled gravitational
systems of physical and numerical interest.  Hamilton-Jacobi methods
have been useful in the past for solving nonlinear problems in cosmology.
In the future, the methods presented in this paper may shed some light on
quantum aspects of the gravitational field.

\noindent
{\bf Acknowledgment}

I would like to thank B. Berger, J. Hartle, J.M. Stewart and W. Unruh
for useful discussions.
This work was supported in part by the Natural Sciences and Research Council of
Canada (NSERC).

\References

\item[] Abramo L R W, Brandenberger R H  and Mukhanov V F 1997
The Energy Momentum Tensor for Cosmological Perturbations
{\it Phys. Rev.} {\bf D56}, 3248-3257

\item[] Berger B K 1998 On the Nature of the Generic Big Bang, in Proceedings
of {\it International Symposium on Frontiers of Fundamental Physics},
Hyderabad, India, Dec. 11-12, 1997

\item[] Berger B K and Moncrief V 1993 Numerical Investigation of
Cosmological Singularities, {\it Phys. Rev.} {\bf D48}, 4676-4687

\item[] Croudace K M,  Parry J,  Salopek D S and  Stewart J M 1994
Applying the Zel'dovich Approximation to General Relativity,
{\it Astrophys. J.} {\bf 423}, 22-32

\item[] DeWitt B S 1967 Quantum Theory of Gravity. I. The Canonical Theory
{\it Phys. Rev.} {\bf 160}, 1113-1148

\item[] Fan Z H and Bardeen J M 1992 Predictions of a Non-Gaussian Model for 
Large Scale Structure, UW-PT-92-11, KEK library scanned preprint: 9207442

\item[] Goldstein H 1981 Classical Mechanics (Addison-Wesley, London)

\item[] Hartle J B 1997 {\it Quantum Cosmology: Problems for the 21st
Century}, in `Physics 2001', Nishinomiya-Yukawa Memorial Symposium on
Physics in the 21st Century: Celebrating the 60th Anniversary of the
Yukawa Meson Theory, Nishinomiya, Hyogo, Japan, 7-8 Nov 1996

\item[]	Hern S D  and Stewart J M 1998 The Gowdy $T^3$ Cosmologies Revisited
{\it Class. Quant. Grav.} {\bf 15}, 1581-1593

\item[] Lanczos C 1970 The Variational Principles of Mechanics
(Fourth Edition, University of Toronto Press, Toronto)

\item[] Linde A, Linde D and Mezhlumian A 1994
{}From the Big Bang to the Theory of a Stationary Universe
{\it Phys. Rev.} {\bf D49}, 1783-1826

\item[] Moscardini L, Borgani S, Coles P, Lucchin F, Matarrese S,
Messina A and Plionis M 1993 Large Scale Angular Correlations
in CDM Models {\it Astrophys. J. Lett.} {\bf 413}, 55

\item[] Parry J, Salopek D S and Stewart J M, 1994
Solving the Hamilton-Jacobi Equation for General Relativity
{\it Phys. Rev. D} {\bf 49}, 2872-81

\item[] Salopek D S 1991
Nonlinear Solutions of Long-Wavelength Gravitational Radiation
{\it Phys. Rev. D} {\bf 43}, 3214-33

\item[]\dash 1992 Cold-Dark-Matter Cosmology with Non-Gaussian Fluctuations
from Inflation {\it Phys. Rev.} {\bf D45}, 1139-57

\item[]\dash 1995 Characteristics of Cosmic Time
{\it Phys. Rev.} {\bf D52}, 5563-75

\item[]\dash 1998  Hamilton-Jacobi Solutions for
Strongly Coupled Gravity and Matter {\it Class. Quantum Grav.} {\bf 15},
1185-1206

\item[] Salopek D S and Bond J R 1990 Nonlinear Evolution of Long-Wavelength
Metric Fluctuations in Inflationary Models {\it Phys. Rev. D} {\bf 42}, 3936-62

\item[]\dash 1991 Stochastic Inflation and Nonlinear Gravity
{\it Phys. Rev. D} {\bf 43}, 1005-31

\item[] Salopek D S and Stewart J M 1992 Hamilton-Jacobi Theory for
General Relativity with Matter Fields {\it Class. Quantum Grav.} {\bf 9},
1943-67

\item[]\dash 1995
Hypersurface-Invariant Approach to Cosmological Perturbations
{\it Phys. Rev.} {\bf D51}, 517-535

\item[] Salopek D S,  Stewart J M and  Croudace K M 1994
The Zel'dovich Approximation and the Relativistic Hamilton-Jacobi Equation
{\it Monthly Notices of the Royal Astronomical Society} {\bf 271},
1005-16

\item[] Saygili K 1997, Hamilton-Jacobi Approach to Pre-Big Bang Cosmology at
Long Wavelengths, CERN preprint, hep-th/9710070

\item[] Starobinski A A (1986) in {Field Theory, Quantum Gravity and
Strings}, proceedings of the Seminar, Meudon and Paris, France,
1984-85, edited by H.T. de Vega and N. Sanchez, Lecture Notes in
Physics, Vol. 246, 107 (Springer-Verlag, New York)

\item[] Unruh W 1998, Cosmological Long Wavelength Perturbations, preprint
gr-qc/9802323

\item[] Veneziano G 1997 Inhomogeneous Pre-Big Bang String Cosmology,
{\it Phys. Lett.} {\bf 406B}, 297-303

\item[] Vilenkin 1983 The Birth of Inflationary Universes
{\it Phys. Rev. D} {\bf 27}, 2848

\item[]\dash 1998 Unambiguous Probabilities in an Eternally Inflating Universe
hep-th/9806185

\endrefs
\end{document}